\documentclass{ieeeaccess}
\usepackage{cite}
\usepackage{amsmath,amssymb,amsfonts}
\usepackage{algorithmic}
\usepackage{graphicx}
\usepackage{textcomp}
\usepackage[caption=false]{subfig}
\def\BibTeX{{\rm B\kern-.05em{\sc i\kern-.025em b}\kern-.08em
    T\kern-.1667em\lower.7ex\hbox{E}\kern-.125emX}}
\begin{document}
\history{Date of publication xxxx 00, 0000, date of current version xxxx 00, 0000.}
\doi{xx.xxxx/xxxx.xxxx.DOI}

\title{Confidence-Aware Learning Assistant}
\author{
    \uppercase{Shoya Ishimaru}\authorrefmark{1},
    \uppercase{Takanori Maruichi}\authorrefmark{2},
    \uppercase{Andreas Dengel}\authorrefmark{1} and
    \uppercase{Koichi Kise}.\authorrefmark{2}
}
\address[1]{German Research Center for Artificial Intelligence (DFKI), Trippstadter Str. 122, 67663, Kaiserslautern, Germany}
\address[2]{Graduate School of Engineering, Osaka Prefecture University, 1-1, Gakuen-cho, Naka-ku, Sakai, Osaka 599-8531, Japan}


\markboth
{Ishimaru \headeretal: Preparation of Papers for IEEE TRANSACTIONS and JOURNALS}
{Ishimaru \headeretal: Preparation of Papers for IEEE TRANSACTIONS and JOURNALS}

\corresp{Corresponding author: Shoya Ishimaru (e-mail: shoya.ishimaru@dfki.de).}

\begin{abstract}
    Not only correctness but also self-confidence play an important role in improving the quality of knowledge. Undesirable situations such as confident incorrect and unconfident correct knowledge prevent learners from revising their knowledge because it is not always easy for them to perceive the situations. To solve this problem, we propose a system that estimates self-confidence while solving multiple-choice questions by eye tracking and gives feedback about which question should be reviewed carefully. We report the results of three studies measuring its effectiveness. (1) On a well-controlled dataset with 10 participants, our approach detected confidence and unconfidence with 81\,\% and 79\,\% average precision. (2) With the help of 20 participants, we observed that correct answer rates of questions were increased by 14\,\% and 17\,\% by giving feedback about correct answers without confidence and incorrect answers with confidence, respectively. (3) We conducted a large-scale data recording in a private school (72 high school students solved 14,302 questions) to investigate effective features and the number of required training samples.
\end{abstract}


\titlepgskip=-15pt

\maketitle

\section{Introduction}
\label{sec:introduction}
\PARstart{Q}{uantified} learning -- sensing learning behaviors for giving effective feedback based on the contexts to each learner -- has high potential in the era of digitalized education~\cite{dengel2016digital}. The appearance of smart sensors equipped on a personal computer, tablet, smartphone, chair, eyeglasses, etc. has enabled researchers to estimate various internal states such as engagement, boredom, tiredness, and self-confidence while learning~\cite{calvo2010affect, warschauer2000technology}. Among these internal states, the importance of self-confidence has especially been investigated in educational research. Self-confidence is a base of metacognitive judgments and the most common paradigm in metacognitive domains ranging from decision-making and reasoning~\cite{ackerman2015, fletcher2012} to perceptual judgments~\cite{fleming2015, peters2015} and memory evaluations~\cite{dunlosky2005, finn2007}. It is a manifestation of metacognitive assessing of one's own knowledge or scholastic ability and affected by proficiency, achievement, cognitive anxiety, and difficulty of a task~\cite{clement1994}. Several studies have reported that a positive increment in self-confidence enhances learners' engagement and performance~\cite{linnenbrink2003, hong2017}.

\begin{figure}[t]
    \centering
    \hspace{-4mm}
    \includegraphics[clip,width=0.96\columnwidth]{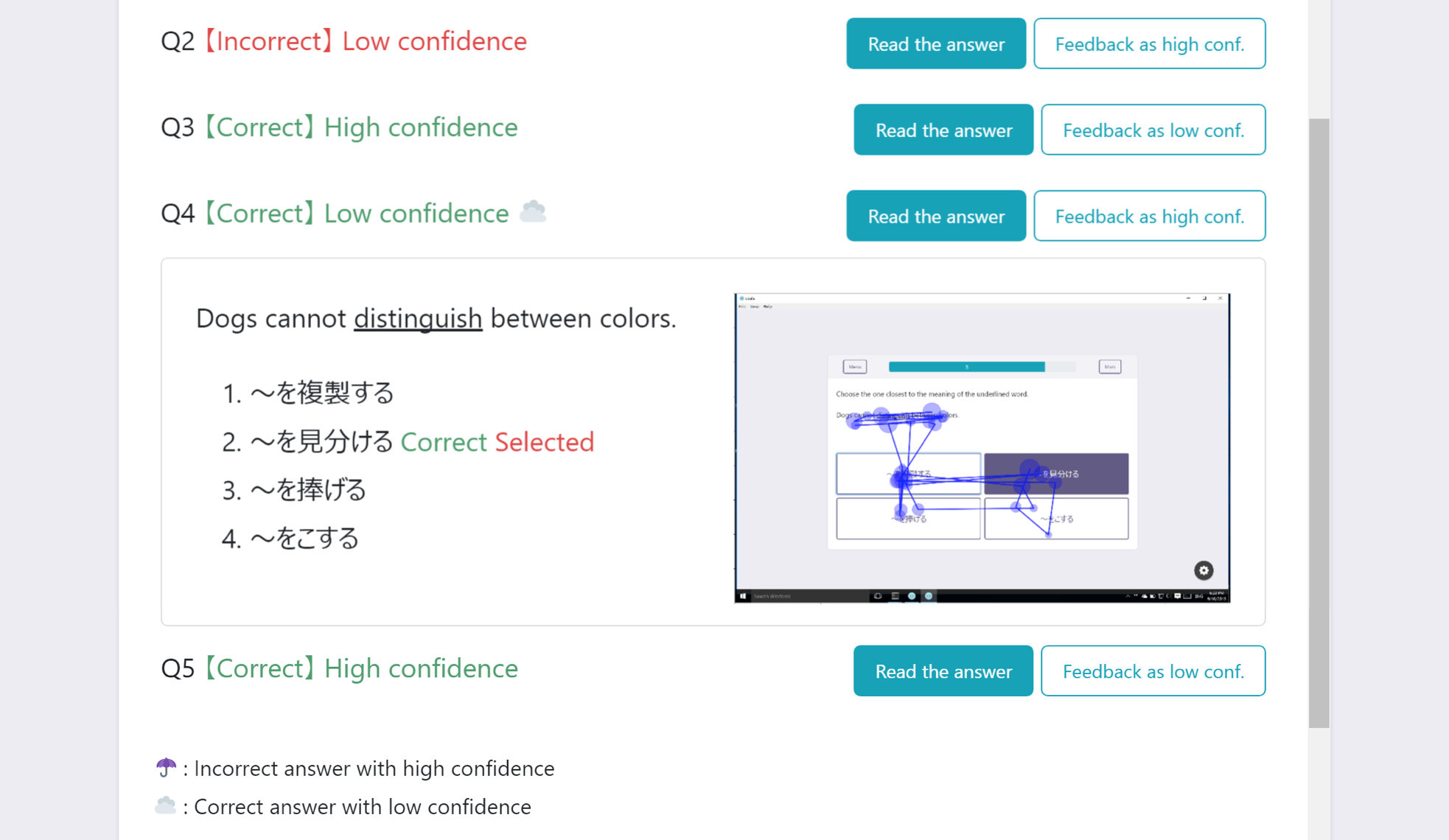}
    \caption{An overview of \textit{Confidence-Aware Learning Assistant (CoALA)}. The system recommends questions which should be checked again on the basis of the correctness and the self-confidence estimated by eye-tracking.}
    \label{fig:overview_report}
\end{figure}

One of the most critical cases where self-confidence plays an important role is on multiple-choice questions (MCQ). MCQ is a type of question that asks selecting the most appropriate choice from given options. Since information obtained from the answer to MCQ is only its correctness, it is hard to distinguish between the cases when a learner answers with confidence or when a learner answers randomly without confidence. We consider that there are four levels of the quality of knowledge, from high to low: correct with confidence, incorrect without confidence, correct without confidence, and incorrect with confidence. In particular, the last two cases are serious. In the case of correct without confidence, the answer will be treated as correct by chance, and the learner loses a chance to acquire correct knowledge. In the case of incorrect with confidence, the wrong knowledge may cause further misunderstandings.

As a solution to such serious cases, we propose a system which estimates self-confidence on MCQ by analyzing eye movements. Based on the estimation output, our system generates a report suggesting which question should be reviewed as shown in Figure~\ref{fig:overview_report}. We define such an intelligent system that adapts learning materials for each learner based on self-confidence as \textit{Confidence-Aware Learning Assistant (CoALA)}. The idea of estimating self-confidence on MCQ by eye tracking has been originally proposed by Yamada \textit{et al.}~\cite{yamada2017}. Compared to their study, we aim for (1) proposing a user-independent approach considering a real scenario and (2) investigating the effectiveness as an end-to-end working system including feedback.

This paper presents our self-confidence estimation algorithm and the results of three studies measuring its effectiveness. In the first study, which involves 10 undergraduate students solving a total of 1,700 questions, we investigated the performance (precision and recall) of the self-confidence estimation. Then we trained a confidence estimator using this dataset and used it for the second study. This is due to a realistic scenario of applying the technology; an estimator is trained with the data from different learners. The second study consists of a pre-test, a review, and a post-test with the help of 20 participants. By comparing the results of the pre-test and the post-test, we evaluated how much our system improves the quality of knowledge in terms of the correctness of the answers and the students' confidence. For the third study, we deployed our system in a private school. During this five-week demonstration, 72 high school students solved 145,489 questions, and 14,302 questions were labeled by themselves. In this wild dataset, we discuss the limitations and future directions of our system. In summary, our contribution in this paper includes:

\begin{itemize}
  \item User-independent gaze-based self-confidence estimation on multiple-choice questions, which detects confidence with 81\,\% and detects unconfidence with 79\,\% average precision, respectively.
  \item End-to-end confidence-based reviewing system, which increases correct answer rates by 14\,\% for unconfident correct answers and 17\,\% for confident incorrect answers compared to a controlled condition.
\end{itemize}

\section{Related Work}
\label{section:related}

\subsection{Eye Tracking for Learning Assistant}
It has been proved that eye-gaze contains linguistic proficiency~\cite{augereau2016} and a degree of self-confidence. For example, the behavior of a student who does not understand the contents of a document is characterized by low reading speed and frequent rereading~\cite{rayner1998}. Thai \textit{et al.} reported that comprehension for a question of a student is appeared in his/her eye movement, for example, rereading of a question~\cite{tsai2012}. Moreover, about the relation between the behavior of eyes and self-confidence, it has been proved that low self-confidence is characterized by a frequent rereading of questions and long gaze on choices~\cite{kojima2012}. Okoso \textit{et al.} have proposed a method of extracting difficult parts of a document for a reader to comprehend and found some effective features~\cite{okoso2015}. Lee \textit{et al.} have proposed to build a virtual tutor to support the learning of a student~\cite{lee2015}. This work demonstrated that eye communications with a virtual tutor enhance the efficiency of learning. Oliver \textit{et al.} have succeeded to estimate the English skill of non-native English speaker from his/her eye behaviors in English test. The contribution of their work is to estimate the skill successfully with a small error by a few documents~\cite{augereau2016}. Yamada \textit{et al.} have tackled the automatic estimation of self-confidence by sensing and analyzing learners' problem solving behaviors through eye movements~\cite{yamada2017}. However, their method works well if the training can be done for each learner with enough amount of data, which may not be realistic. In other learning subjects, for instance, Ishimaru \textit{et al.} have investigated reading behaviors of students on a textbook in Physics~\cite{ishimaru2018augmented}. They have proposed Areas of Interests based and subsequence based approaches to predict expertise.

\subsection{Other Sensing Modalities}
Some researchers have measured the attention of students in learning by Electroencephalogram (EEG) and investigated a correlation between attention and self-efficacy, which refers to the level of confidence of an individual with regard to their ability at task execution)~\cite{sun2017,hsia2016}. Though a method using EEG can be a solution to estimate self-confidence, the device disturbs a user engaging in a task for the reason that it is always attached to his head. On this point, the eye tracker is preferable because it can be attached to a display.

There has been a growing interest in the study of the relation between cognitive performance and the autonomic nervous system (ANS). The activity of ANS can be also measured by heart rate variability (HRV)~\cite{luque2013cognitive}, Electrodermal activity (EDA)~\cite{brishtel2018assessing, critchley2002electrodermal} and so on. We did not utilize these approaches in this work because we had received comments from students in the private school that wearing sensors while studying requires a high physical workload. If we can record the precise physiological signal with remote sensing, we consider integrating it. For instance, the nose temperature, which can be measured by a commercial infrared thermography camera, can be a nice candidate~\cite{ishimaru2017cognitive}. Abdelrahman \textit{et al.} have recorded nose and forehead temperatures under different task difficulties and found significant changes~\cite{abdelrahman2017cognitive}.

Although mobile eye trackers appeared, there is still a strong gap between controlled behaviors in the laboratory and natural behaviors in the wild. One of the critical issues in this research field is how we can conduct experiments in natural settings for proposing robust methods. Towards this objective, several researchers have conducted long-term and large-scale experiments (e.g., over 80 hours of recording with a mobile eye tracker~\cite{steil2015discovery} and 780 hours of recording with a commercial Electrooculography glasses~\cite{ishimaru2017towards}). Our work is in this context and has evaluated real learning behaviors.

\subsection{Role of Self-Confidence in Learning}
Several studies have mentioned correlations between self-confidence or other cognitive states and behaviors of people in specific tasks including achievement test of learning~\cite{jersakova2017},
cognitive test~\cite{kleitman2011},
and cooking~\cite{pooler2017}.

According to work by Forbes-Riley \textit{et al.}, adapting a user's self-confidence into the computer tutoring system improves performance on learning efficiency and a user's satisfaction~\cite{forbes2009adapting}. Kleitman \textit{et al.} reported that a high level of self-confidence predicted high grades for primary school children~\cite{kleitman2012metacognitive}. Indeed, students who have self-confidence awareness tend to be recognized in their performances, which develops their level of self-confidence again. This positive feedback loop motivates students to learn by themselves. In another study, Stankov \textit{et al.} showed that self-confidence can be used to identify misconceptions~\cite{stankov2012confidence}. The misconception occurs when a learner feels confident with the knowledge and thinks that he/she is answering correctly but actually gives an incorrect answer.

Roderer \textit{et al.} have gathered participants of several ages and have found a correlation between the self-confidence of participants and their age. Junior participants have tended to get higher self-confidence than senior participants~\cite{roderer2014}. In contrast to this research, we gathered participants of almost the same age so as to investigate self-confidence with only information in answering. The researches referred above, however, only have proposed the importance of self-confidence. On the other hand, our work is not only to find correlations but also to estimate self-confidence for practical applications.

\subsection{Position of our work}

Most of the previous work focuses only on the scientific investigation about the importance of self-confidence. Only a limited number of research trials have tackled the automatic estimation of confidence. Moreover, the use of estimated confidence to improve the quality of knowledge has not been well attempted in the past. We consider that this is due to the following two limitations.

\textbf{No general estimation} -- It is difficult to establish self-confidence estimation which is independent of environments, subjects, and learners. In other words, estimation methods may work well under a specific environment, for specific subjects and learners, but may not if those conditions are no longer satisfied. In the latter case, the estimation is unstable and less reliable. The important research question here is whether such estimation is still effective in improving the quality of knowledge.

\textbf{No end-to-end system} -- Effectiveness should be evaluated as an end-to-end system including sensing, estimation, and feedback. It is often the case that parts work well but the system built by connecting them does not. Unfortunately, most of the previous work focuses on parts, and little has considered the end-to-end scenario. If the goal is to build a system capable of improving the quality of knowledge of learners, this standpoint is mandatory.

In summary, our work's main aim is to evaluate all the methods of sensing, estimation, and feedback not independently but as an end-to-end system to prove that it can improve the quality of knowledge.


\begin{figure}[t]
    \centering
    \subfloat[Answer with confidence]{
      \includegraphics[clip,width=0.48\hsize]{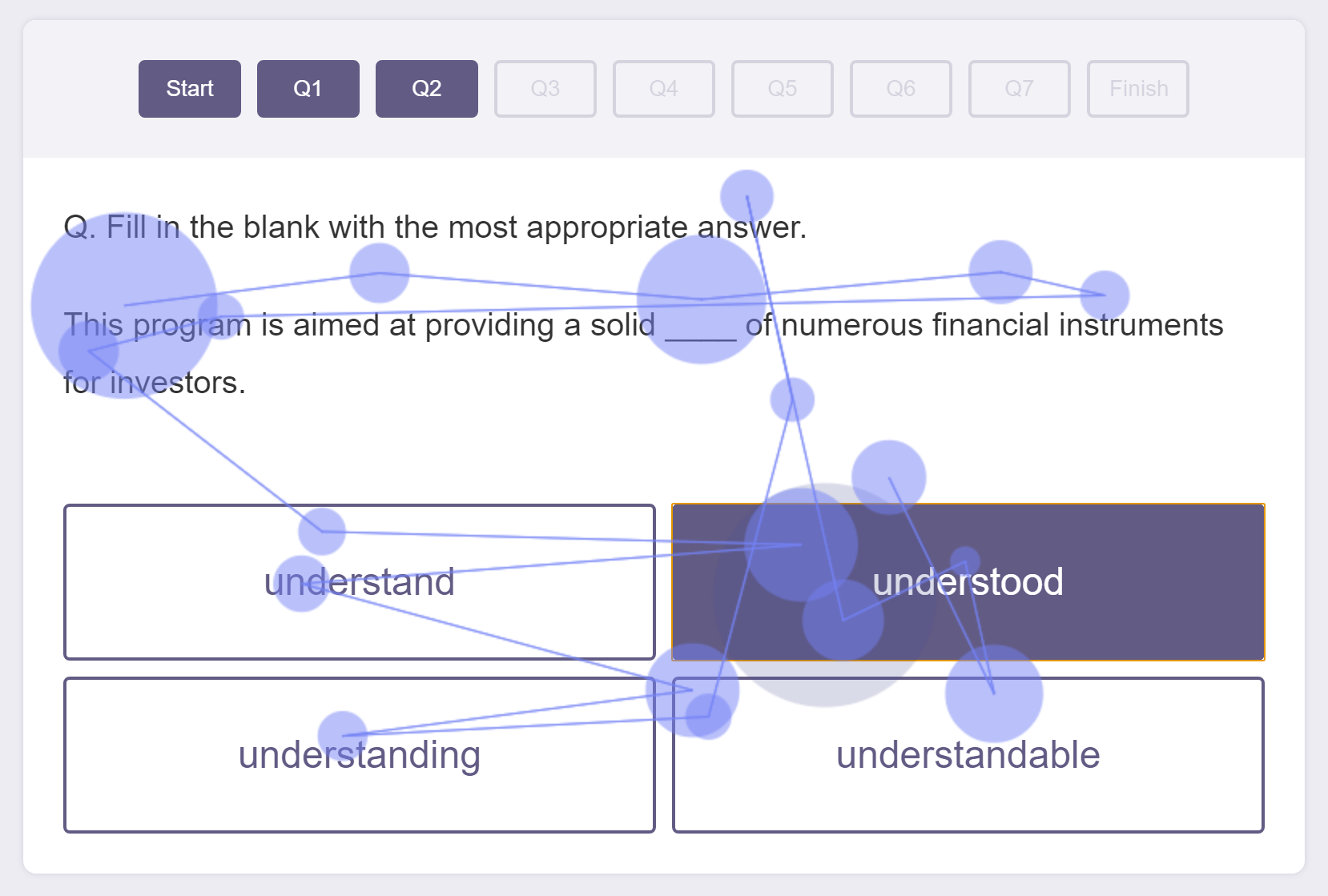}
      \label{fig:feedback_corr_uncf}
    }
    \subfloat[Answer without confidence]{
      \includegraphics[clip,width=0.48\hsize]{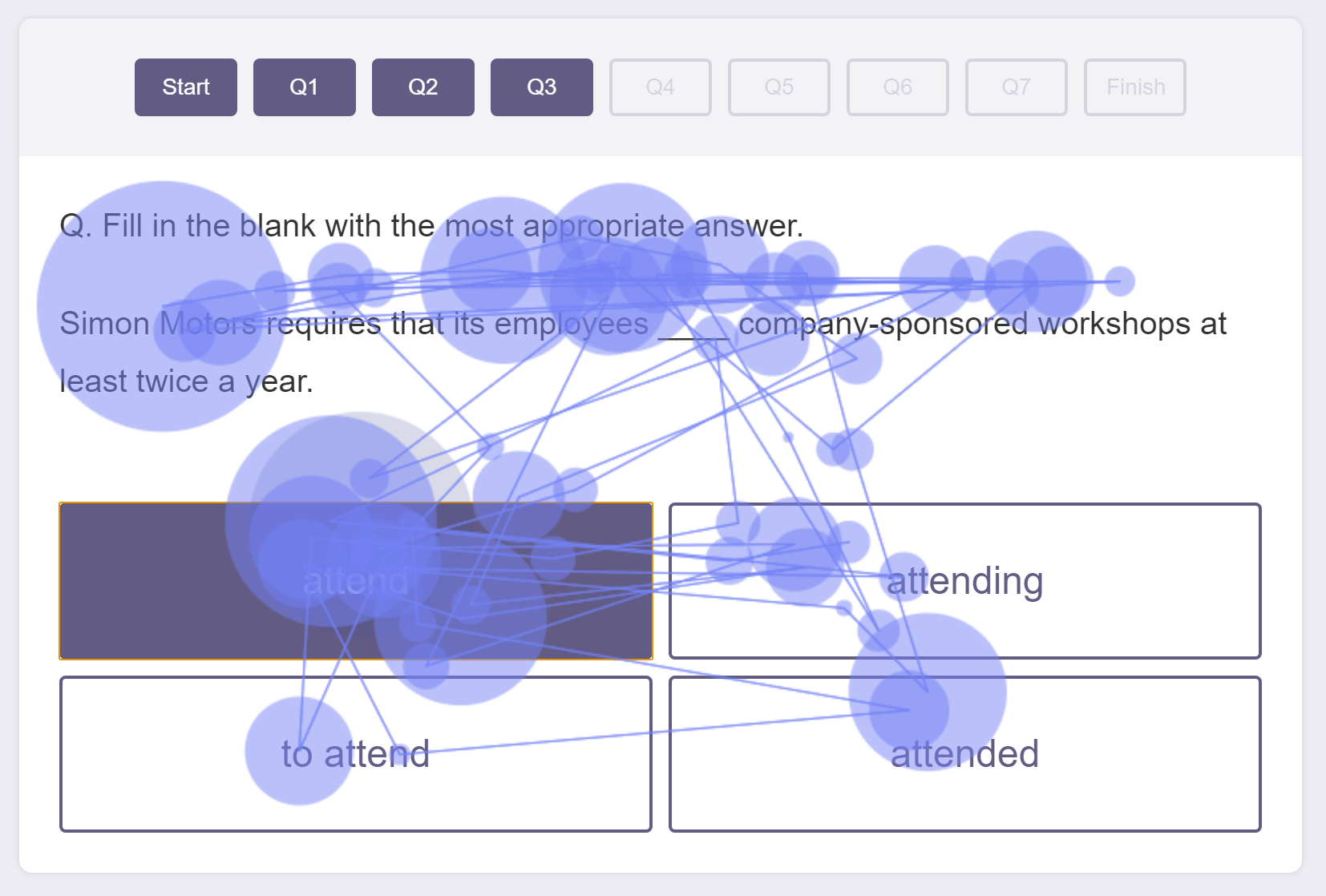}
      \label{fig:feedback_incr_conf}
    }
    \caption{Examples of eye gaze on multiple-choice questions}
    \label{fig:example}
  \end{figure}

\section{Proposed Method}
The processing in our system consists of the following four steps: data recording, feature calculation, feature selection, classification, and feedback.

\subsection{Data Recording}
The eye gaze of a user is recorded by a remote eye tracker attached at the bottom of a display. The output of the eye tracker includes coordinates of the gaze on display and their timestamps. Figure~\ref{fig:example} shows the difference of eye gaze while solving MCQ with and without confidence. This preliminary observation indicates that confusion of choices appears in eye gaze as the transition between choices. The circle in the figure is the position the user is looking at, which is visualized as a demonstration and calibration purpose, invisible while solving questions.

Eye movements are composed of two events: fixations and saccades. A fixation indicates an event when the gaze pauses at a certain position over a certain period, usually minimum 100\,ms. A rapid movement between fixations is called a saccade. We classify raw gaze into fixations and saccades by using an algorithm proposed by Buscher \textit{et al.}~\cite{buscher2008}. A blink -- rapid closing of the eyelid -- is not analyzed in our method because the time required to solve one question (10 - 60 sec.) is too short of calculating statistical features. In addition, a smooth pursuit occurs when a person tracks a moving object with slow speed. But this metric is not considered in this method because all information on a display is fixed.

\subsection{Feature Calculation}

We define Areas of Interest (AOIs) as rectangles covering a question and each choice in order to recognize deep behaviors (e.g., a ratio of reading-times on a question and choices, a process of the decision with comparisons of choices, etc.) Fixations and saccades are automatically associated with the corresponding AOIs in this step. Then we extract 30 features shown in Table~\ref{tab:features}. Features 1--14 are related to fixations, and Features 15--28 are about saccades. We also use the reading-time and the correctness of the answer as features.

\begin{table}[t]
    \begin{center}
    \caption{The list of the features}
        \label{tab:features}
        \begin{tabular}{cl} \hline
            No.   & Feature \\ \hline
            1-2   & Fixation \{count, ratio\} on Choices \\
            3-4   & Fixation \{count, ratio\} on Question \\
            5-8   & \{Sum, mean, max, min\} of fixation durations on Choices \\
            9-12  & \{Sum, mean, max, min\} of fixation durations on Question \\
            13-14 & Variance of \{x, y\} coordinate of fixations \\
            15-16 & \{Sum, mean\} of saccade length \\
            17-20 & Saccade count \{all, on Question, between Choices, \\
                  & between Question and Choices\} \\
            21-24 & \{Sum, mean, max, min\} of saccade durations \\
            25-28 & \{Sum, mean, max, min\} of saccade speeds \\
            29    & Reading-time \\
            30    & Correctness of the answer \\ \hline
        \end{tabular}
    \end{center}
\end{table}

\subsection{Feature Selection}
We select effective features from the above 30 candidates because increasing the number of features does not always increase classification performance. We utilize the following simple hill-climbing strategy called forward stepwise. Firstly, we create a subset of features. The subset is empty at the initial state. Then we calculate average precision scores of estimations using each feature and insert one with the best feature to the subset. Performances of estimations with features in the subset and one new feature are calculated, and keep the best combination again. These processes are repeated as long as the new subset performs better than the old one. Two-fold cross-validation is used for this feature selection. Note that this step has proceeded only while training a classifier with training samples. Then selected features are used to classify unknown samples.

\subsection{Classification}
We estimate the self-confidence of answers by a Support Vector Machine (SVM) with the selected features. The Radial Basis Function (RBF) kernel with penalty parameters $C = 1$ and $\gamma = 0.125$ were selected experimentally and are used for the SVM. As a preliminary study, we tested other machine learning techniques including Random Forest, and found that SVM performs the best overall in our classification task.

\subsection{Feedback to a Learner}
By combining the correctness and the estimated confidence, answers of a learner are categorized into four groups: correct with confidence, correct without confidence, incorrect with confidence, and incorrect without confidence. As shown in Figure~\ref{fig:overview_report}, the system highlights questions that should be specially reviewed. A learner can claim if the output is wrong. Then the data are stored to personalize the upcoming estimation.

\section{Study 1: Evaluation of Self-Confidence Estimation}

In this first study, we evaluate the performance of the self-confidence estimation. We involved 10 participants with the same background in creating a well-designed dataset. This section explains the procedures and the results.

\subsection{Experimental Design}
We invited 10 participants (male: 5, female: 5) to our laboratory for solving 170 MCQ about English vocabularies and grammars. All the participants were first-year Japanese undergraduate students. We utilized Tobii 4C remote 90\,Hz eye tracker for this data recording. Note that an upgrade key provided by Tobii was applied to use this device for scientific purposes. Participants answered the most appropriate word for a blank in a question from choices. After answering each question, they answered a survey \textit{``Do you have a confidence in your decision?''} with \textit{Yes} or \textit{No}. Answers to this questionnaire were used as ground truth labels (referred to as \textit{true confidence} in this paper). We applied the random oversampling in \textit{imbalanced-learn} to create a balanced dataset.


\subsection{Self-Confidence Estimation Performance}

\begin{figure}[t]
    \centering
    \hspace{-4mm}
    \includegraphics[clip,width=0.96\hsize]{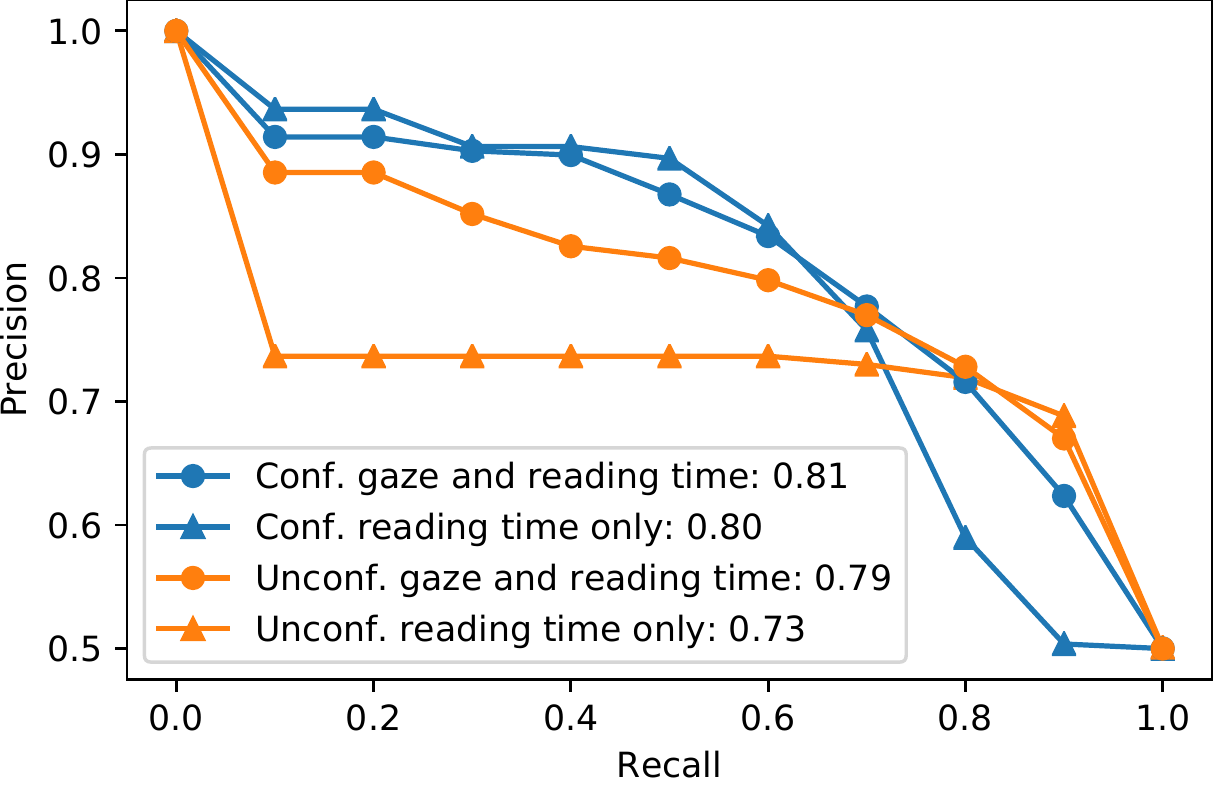}
    \caption{11-point precision-recall curves and average precisions of self-confidence estimations}
    \label{fig:estimation_estimation}
\end{figure}

Figure~\ref{fig:estimation_estimation} shows 11-point precision-recall curve of the confidence detection and unconfidence detection among all participants. This result indicates that our confidence estimation performs accurate enough, relatively better in confidence detection compared to unconfidence detection (average precisions: 81\,\% and 79\,\%). Since the labels of confidence were balanced, the chance ratio of the estimation is 50\,\%. Selected features from this recording were as follows: \textit{f5: sum of fixation durations on choices}, \textit{f13: variance of x coordinate of fixations}, \textit{f19: the number of saccades between choice areas}, \textit{f21: sum of saccade durations}, and \textit{f29: reading-time}. Since some of the selected features are correlated with each other, one feature, i.e., only reading-time might be able enough to classify confident and unconfident. However, the eye gaze feature improves the performance, in particular for the unconfident detection.

\subsection{Observation of Misclassifications}

\begin{figure}[t]
    \centering
    \subfloat[True conf. estimated as conf.]{
        \includegraphics[clip,width=0.48\columnwidth]{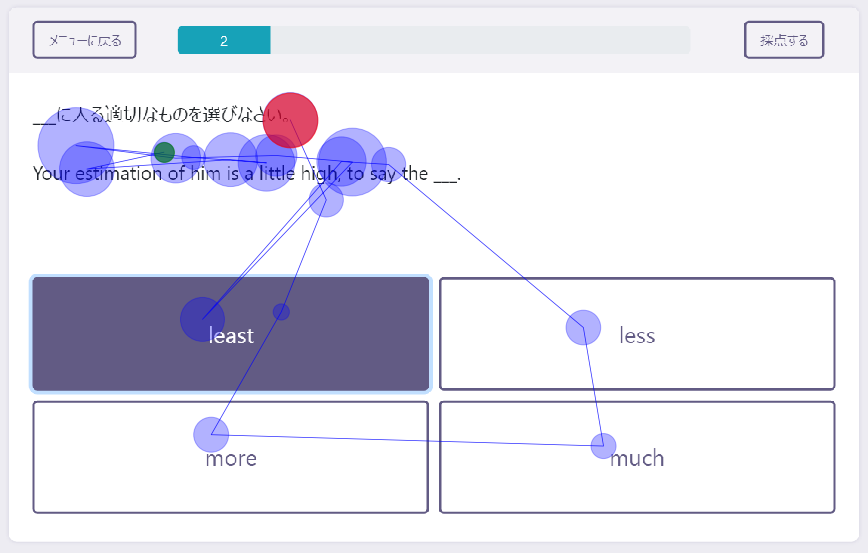}
        \label{fig:lc_ec}
    }
    \subfloat[True unconf. estimated as conf.]{
        \includegraphics[clip,width=0.48\columnwidth]{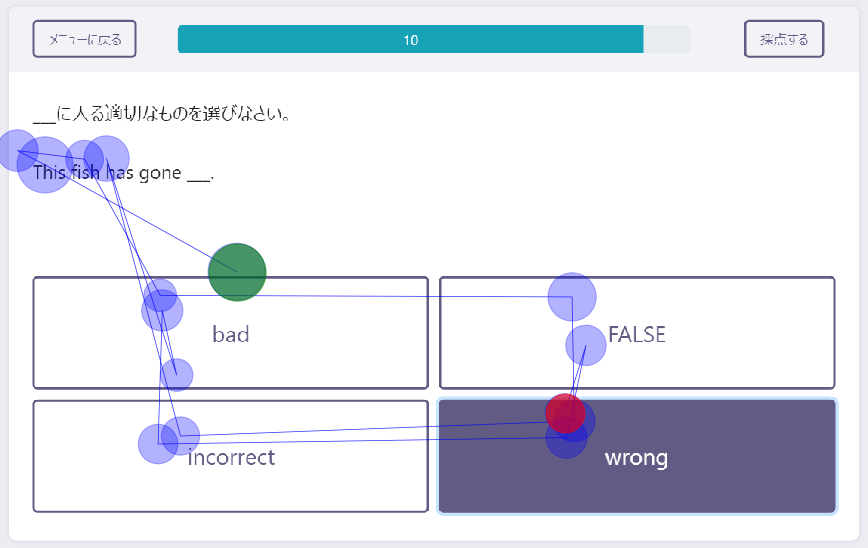}
        \label{fig:li_ec}
    }
    \newline
    \subfloat[True conf. estimated as unconf.]{
        \includegraphics[clip,width=0.48\columnwidth]{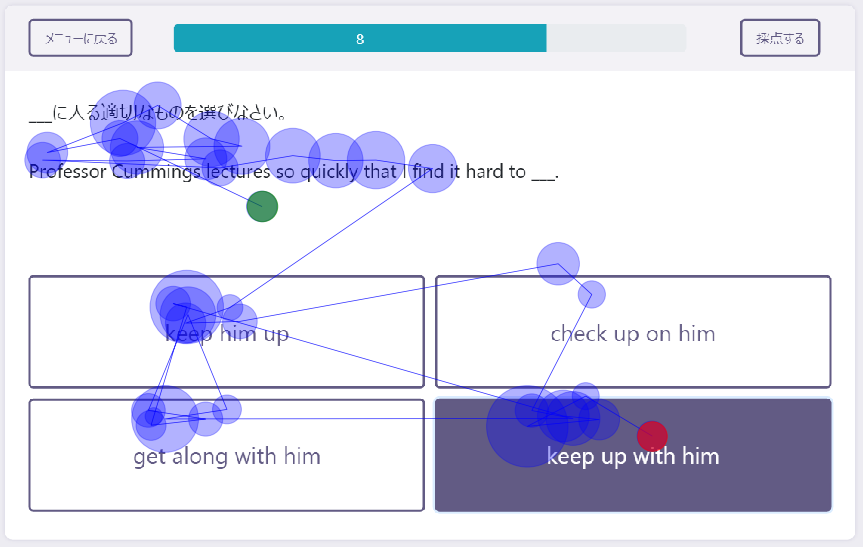}
        \label{fig:lc_ei}
    }
    \subfloat[True uncof. estimated as unconf.]{
        \includegraphics[clip,width=0.48\columnwidth]{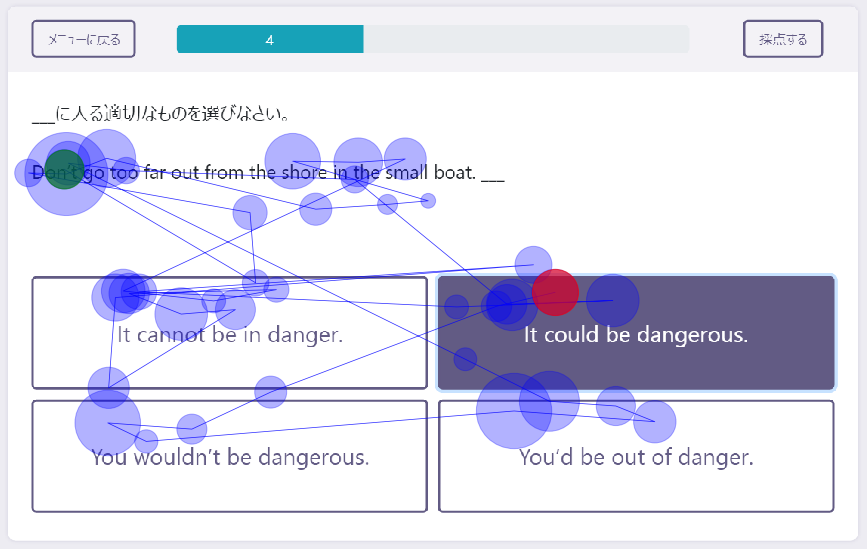}
        \label{fig:li_ei}
    }
    \caption{Examples of eye gaze on each classification result}
\label{fig:gaze}
\end{figure}

We describe a difference in the eye gaze between the case a participant answered with confidence and without confidence. We display on Figure~\ref{fig:gaze} some examples of the estimation results. The circles represent the fixations, and the diameter of the circle is proportional to fixation duration. Hence the longer a participant looked at a point, the larger the diameter of the fixation is. The lines between circles represent the saccades.

Figure~\ref{fig:gaze}~\subref{fig:lc_ec} is an example of the eye gaze of a participant answered with confidence. Figure~\ref{fig:gaze}~\subref{fig:li_ei} is an example of the eye gaze which a participant answered without confidence. We can find that the confidence in answering is characterized by the fewer eye movements and smaller diameter of the fixations, on the other hand, the unconfidence is characterized by the complex eye movements and the longer fixation durations.

In Figure~\ref{fig:gaze}~\subref{fig:li_ec}, a participant answered without confidence, but the classifier estimated as he answered with confidence. We assume that he gave up to answer correctly to this question because he did not have the necessary knowledge. In such a case, we can find that the number of fixations is small, and the participant took a short time to answer this question. These characteristics are common to Figure~\ref{fig:gaze}~\subref{fig:lc_ec}, which represents a confident decision. Therefore the classifier estimated as a confident decision.

In Figure~\ref{fig:gaze}~\subref{fig:lc_ei}, a participant answered with confidence, but the classifier estimated as he answered without confidence. We assume that this participant decided his answer carefully by eliminating irrelevant choices one by one. In such a case, we find more fixations and frequent transitions of eyes between rectangles. This characteristic is common to Figure~\ref{fig:gaze}~\subref{fig:li_ei}, which represents an unconfident answer.

\section{Study 2: Evaluation of Self-Confidence Based Feedback}

To evaluate the effectiveness of feedback based on self-confidence, we utilized the classifier of the first recording and prepared the end-to-end review feedback system as the second study. This section explains the details of the experiment and answers to the following our research hypotheses.

\begin{figure}[t]
    \begin{center}
    \includegraphics[clip,width=0.96\hsize]{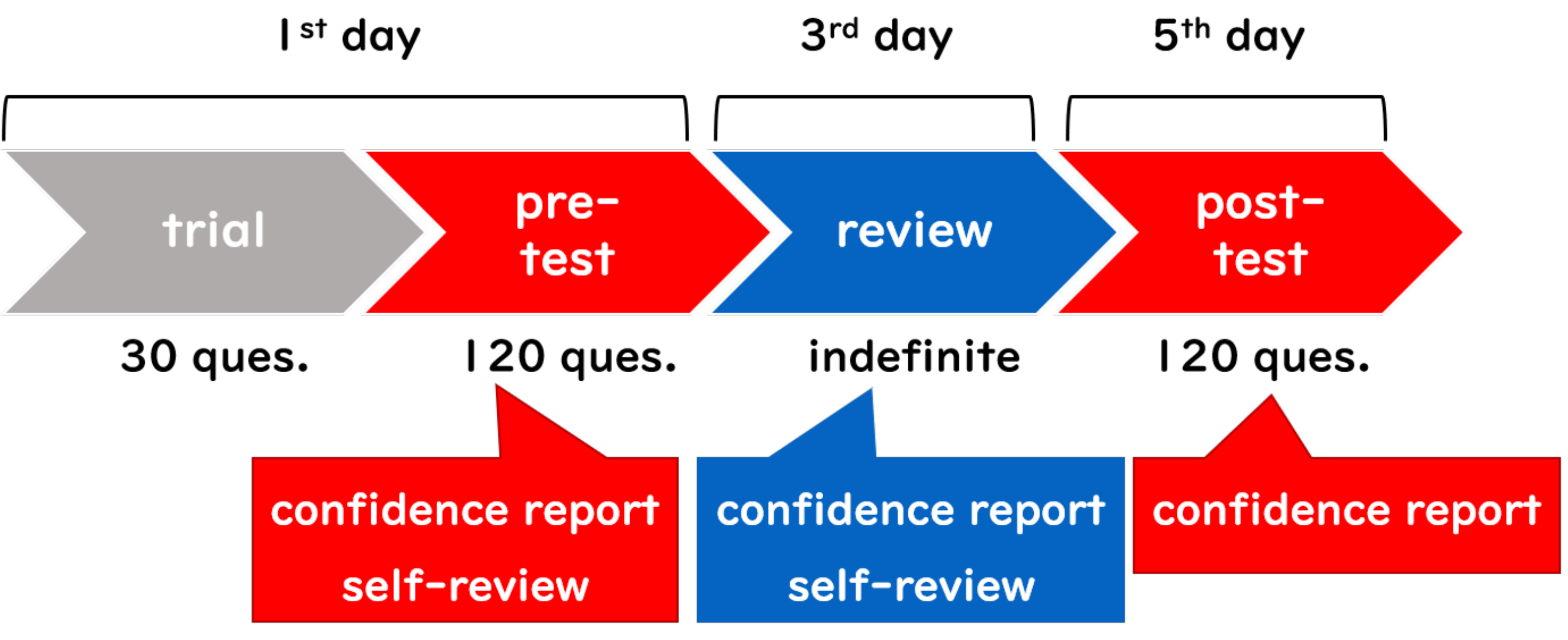}
    \caption{The procedure of the feedback study.}
    \label{fig:feedback_procedures}
    \end{center}
\end{figure}

\begin{itemize}
    \item \textbf{RH1} -- Questions answered correctly without confidence (vague knowledge) tend to be forgotten compared to knowledge with confidence.
    \item \textbf{RH2} -- Questions answered incorrectly with confidence (misunderstandings) tend to be mistaken again compared to wrong knowledge without confidence.
    \item \textbf{RH3} -- Estimating self-confidence from learning behaviors and giving feedback (e.g., adding questions to a review list, highlighting them while reviewing) avoids such scenarios.
\end{itemize}

\subsection{Experimental Design}

We employed 20 participants (undergraduate and graduate school students, age: 18--25, male: 14, female: 6) and monitoring the transition of their performance. For questions, we prepared three levels of MCQ about English grammar: Level 1 (easy, 170 questions), Level 2 (normal, 290 questions), and Level 3 (hard, 160 questions). Each question requires the most appropriate word for a blank in a question from four choices. Eye movements on the questions were recorded by Tobii 4C remote 90\, Hz eye tracker with an upgrade key. Figure~\ref{fig:feedback_procedures} shows the experimental procedure. We invited participants for three days and asked for the following tasks. One-day breaks were inserted between each task-day. Participants who completed tasks received 5,000 JPY.

\textbf{Trial (the first day)} --
Each participant solved 10 questions with the three levels as a trial. Two reasons were behind this trial: getting participants used to the MCQ interface and selecting an appropriate degree of difficulty. If the questions are too easy or too difficult, the dataset will be unbalanced, and we cannot show any transition of their performance. Based on the results, we selected a suitable level whose correct answer rate is closest to 50\,\%.

\textbf{Pre-Test (the first day)} --
After choosing the suitable level, each participant answered 120 questions of the selected level and reported his/her self-confidence after each decision. Besides, the result page (see Figure~\ref{fig:overview_report}) with correctness and estimated self-confidence based on the training dataset appears after answering every 10 questions. We instructed participants to press ``Read the answer'' button for self-review except for the questions correctly answered with confidence.

\begin{figure}[t]
    \begin{center}
    \includegraphics[clip,width=1.0\hsize]{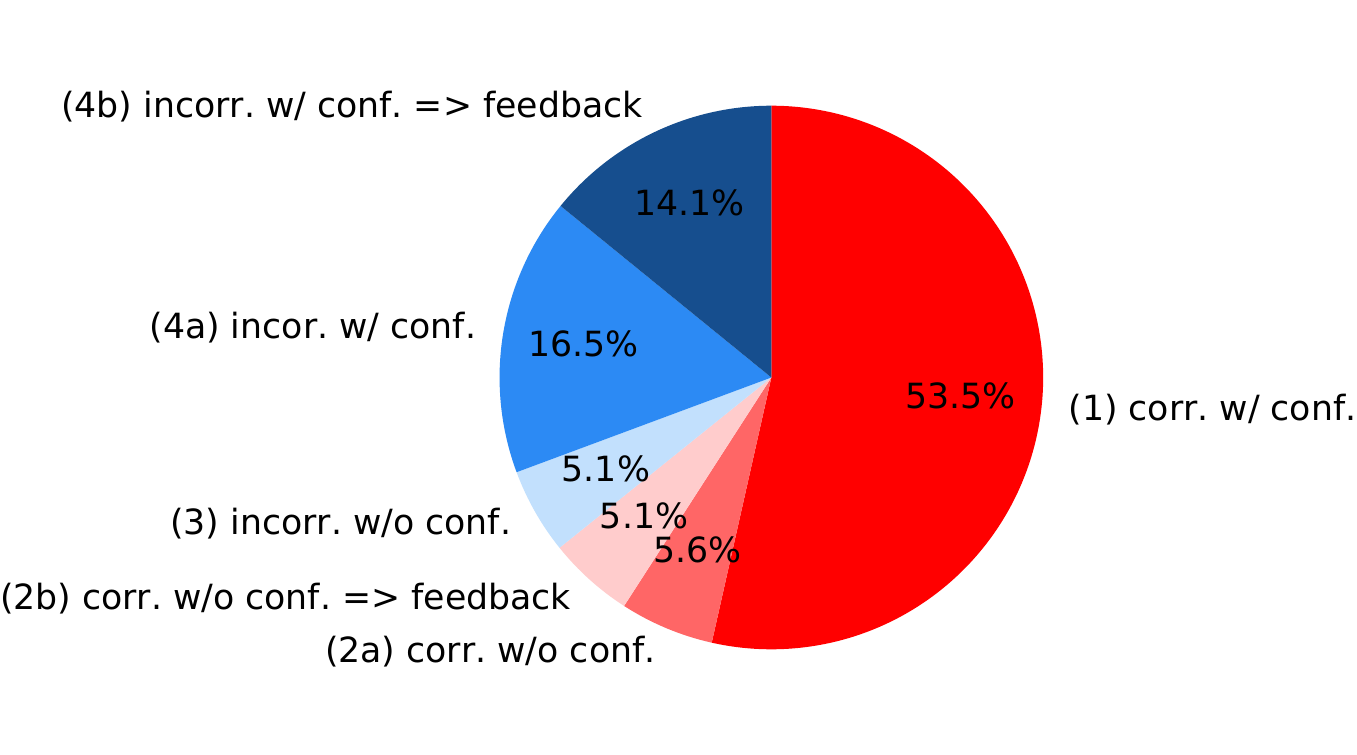}
    \caption{The distribution of questions in the feedback study.}
    \label{fig:feedback_distribution}
    \end{center}
\end{figure}

We recorded 2,075 answers in total. Based on the correctness and estimated self-confidence, we categorized them into four groups: (1) correct with confidence, (2) correct without confidence, (3) incorrect without confidence, and (4) incorrect with confidence. The role of our system is to identify (2) and (4) for suggesting a learner review them. In order to evaluate the effectiveness of the system, we gave feedback to half of (2) and (4) (see Figure~\ref{fig:feedback_distribution}). In following, the \textit{without feedback} samples are called as controlled groups (2a) and (4a), and the \textit{with feedback} samples are defined as experiments groups (2b) and (4b).

\textbf{Review (the third day)} --
Participants answered review questions generated based on the first day's feedback. Wrong answers (3) and (4) were inserted into the list of the review. In addition, we added (2b) to the list. During the review, (2b) and (4b) were emphasized on the question page. After solving each question, each participant reported his/her self-confidence. The result page with correctness and estimated self-confidence was shown for every 10 questions. We asked participants to press ``Read the answer'' button again for self-review, except for the questions correctly answered with confidence. The order of questions and choices was shuffled from the pre-test.

\textbf{Post-Test (the fifth day)} --
Participants solved the same 120 questions as the pre-test. They reported confidences on decisions for each question and checked the result page every 10 questions as same as the first and the third day. The order of questions and choices was shuffled from the review.

\subsection{Importance of Self-Confidence Estimation}
\begin{figure}[t]
    \centering
        \subfloat[Correct at the pre-test]{
        \includegraphics[clip,width=0.5\hsize]{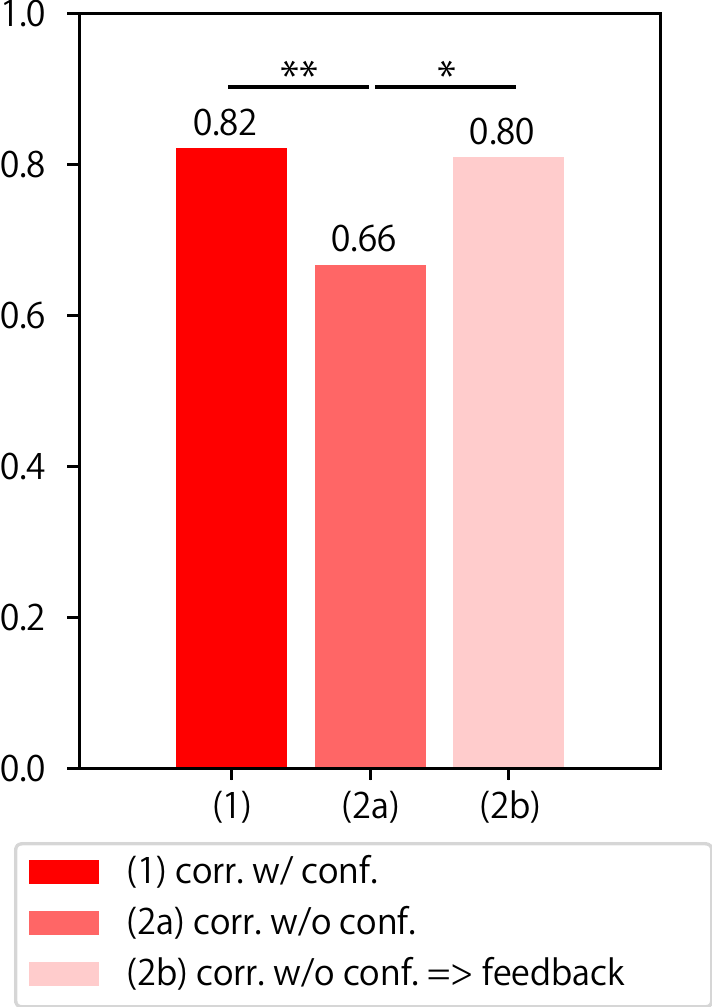}
        \label{fig:corr_corr_all}
        }
        \subfloat[Incorrect at the pre-test]{
        \includegraphics[clip,width=0.5\hsize]{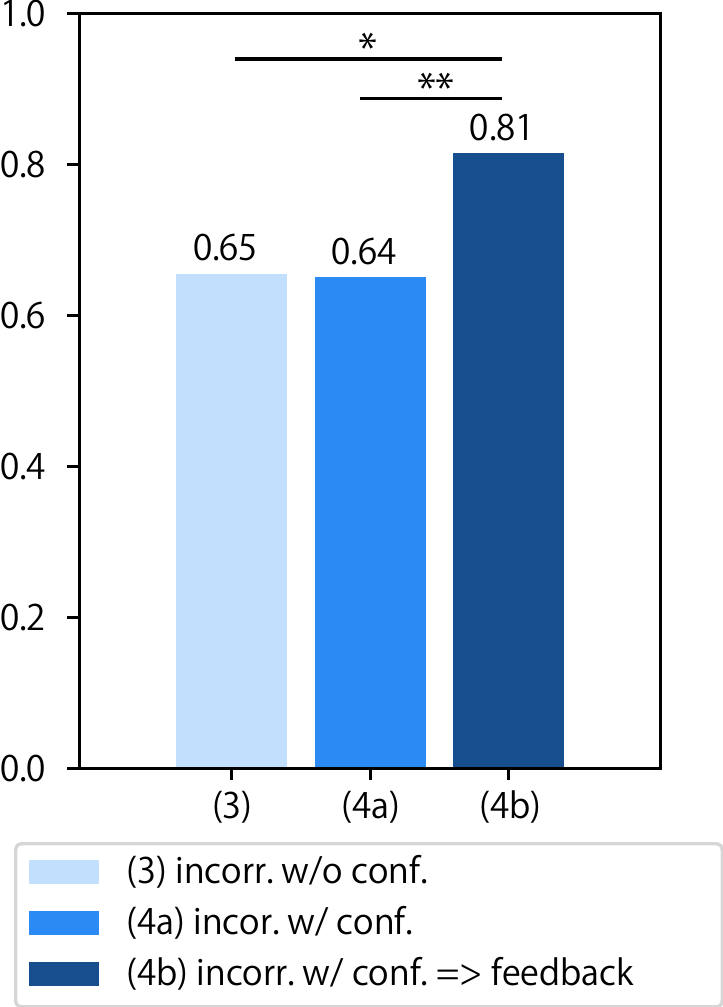}
        \label{fig:incr_corr_all}
        }
    \caption{The mean of correct answer rates among 20 participants at the post-test. The symbol ** and * indicates \textit{p}<0.01 and \textit{p}<0.05, respectively.}
    \label{fig:corr_all}
\end{figure}

Figure~\ref{fig:corr_all} shows results of the effect of our review feedback. For this investigation, we divided all questions into two groups: answered correctly or incorrectly at the pre-test. Then we compared their correctness at the post-test under each condition.

If a participant answered correctly at the pre-test, he/she should be able to select the right choices again when he/she is asked the same questions. However, some answers were wrong at the post-test. Figure~\ref{fig:corr_all}~\subref{fig:corr_corr_all} reports how many questions were forgotten. As a result, the ratio of (2a) correct answers without confidence was dropped 16\,\% compared to (1) answers with confidence ($p<0.01$ evaluated by Welch's t-test). In other words, answers without confidence tend to be forgotten in the near future, and therefore they should be included in the review list. We observed that questions answered without confidence could not always be answered correctly if they are asked again (\textbf{RH1} is true). There is not much difference in the correctness of post-test between wrong answers with and without confidence (\textbf{RH2} is not always true).

\subsection{Effect of Feedback}
Figure~\ref{fig:corr_all} also shows that feedback succeeded to improve the mean correct answer rate. The performance of the experiment groups was 14\,\% higher than the performance of the controlled group ($p<0.05$) for the feedback about correct and unconfident questions (see Figure~\ref{fig:corr_all}~\subref{fig:corr_corr_all}), and was 17\,\% higher for incorrect and confident questions ($p<0.01$; see Figure~\ref{fig:corr_all}~\subref{fig:incr_corr_all}). Highlighting questions that were answered incorrectly with confidence could increase the probability of maintaining the correct answers in mind (\textbf{RH3} is true).



\subsection{Quality of Knowledge}
\begin{figure}[t]
    \includegraphics[clip,width=0.98\hsize]{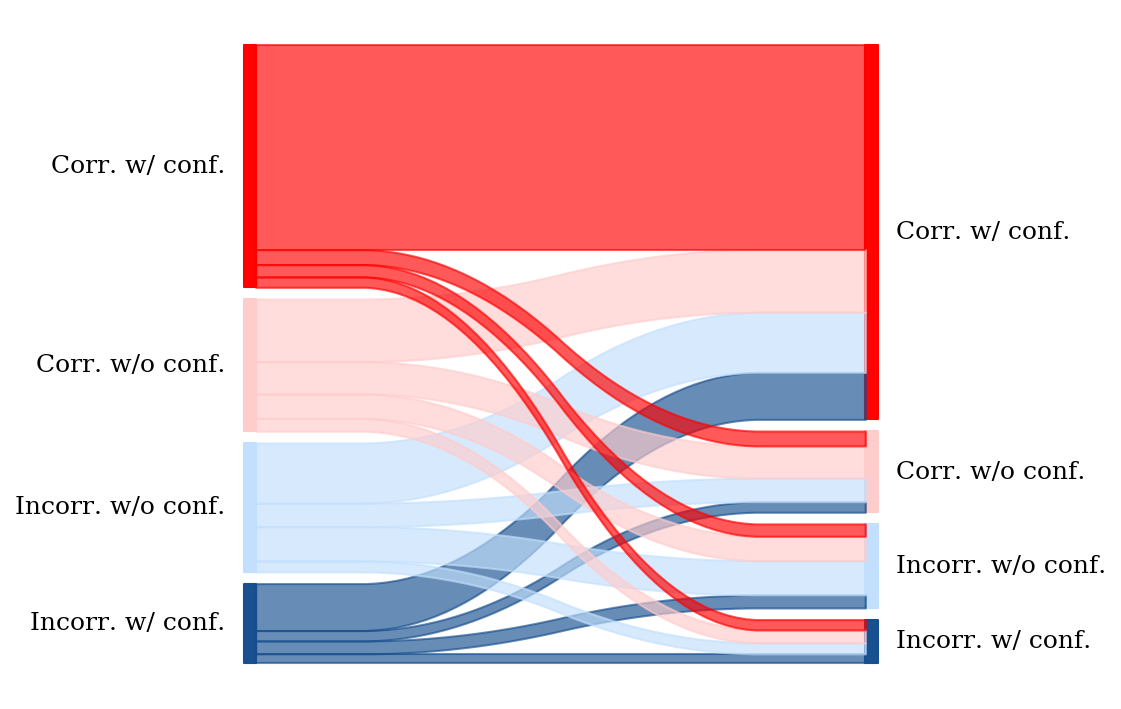}
    \caption{Transitions of currentnesses and estimated confidences before (left) and after (right) the review.}
    \label{fig:feedback_sankey}
\end{figure}

Let us show how the quality of knowledge changes by the feedback with the estimated confidence. Figure~\ref{fig:feedback_sankey} represents transitions of levels: correctness and reported confidence between the pre-test and post-test. Controlled groups (randomly selected no feedback samples) are not included in this chart. After the review, the number of correct answers with confidence was increased compared to the other three groups. In addition, an interesting finding from this chart is that participants were able to assess their state of knowledge better after the review. A lot of correct with unconfident answers were changed to correct with confidence answers. And percentages of correct with unconfident answers and incorrect with confident answers were decreased. From the result mentioned above, the feedback is effective in improving the quality of knowledge.

\section{Study 3: Deployment in the Wild}

This paper has demonstrated how does CoaLA estimate self-confidence and how much does it improve learning performances. However, it is a common affair that unexpected problems in the laboratory condition happen in the wild condition. In this section, we report findings from a deployment in the real classroom environment.

\subsection{Experimental Design}

We have collaborated with a private school and deployed our system in the school. Students solved MCQ about vocabularies in English on the system. Then they printed out a list of words involving incorrect answers and correct answers with low self-confidence. The questions were prepared by the private school. The main purpose of this deployment is not to record data but to demonstrate the system in the real environment. Therefore, unlike the previous two studies, we did not prevent students' natural behaviors. Calibration of an eye tracker was performed once before a student starts using the system. We asked the self-confidence of the decision (ground truth labels) once every five questions. Each student has their own username in order to track who solved which question with or without confidence. The number of solved questions depends on the students. We utilized Tobii 4C remote 90\,Hz eye tracker with an upgrade key. The duration of this demonstration was around five weeks. 83 students used our system, and we collected 145,489 solving behaviors in total. We evaluated our proposed self-confidence estimation on this dataset with leave-one-participant-out cross-validation.

\subsection{Pre-processing}

Since real recordings included many noisy behaviors, the following filterers were applied to obtain a reliable dataset. (1) We analyzed labeled data in this study. (2) Data with invalid usernames (e.g., \textit{guest}) are filtered out. (3) Data with only a few eye gaze (a ratio of valid gaze coordinates is less than 80\,\% of one recording) are also ignored. Finally, the wild dataset consists of 14,302 valid samples from 72 students.

In a real learning scenario, we are not able to ask students to calibrate an eye tracker many times. They frequently move ahead and change a seat position. Therefore eye gaze in the wild dataset was not precise compared to data in the laboratory. It causes problems in our feature calculation because AOIs are predefined as absolute coordinates on display. However, an interesting finding from scan path images is that a relative positional relationship between gazes on a question and choices is still correct even if they are shifted. In order to solve this issue, we decided to define AOIs with a new approach. From all fixations in one recording, we calculate the maximum and the minimum $x$ and $y$ coordinates. Then AOIs are defined on the basis of relative positions in this space. In our question format, an area of question is the 34\,\% top part of the space, and areas of questions are divided into a cross of the remaining 66\,\% bottom part.


\begin{figure}[t]
    \centering
    \hspace{-4mm}
    \includegraphics[clip,width=0.96\hsize]{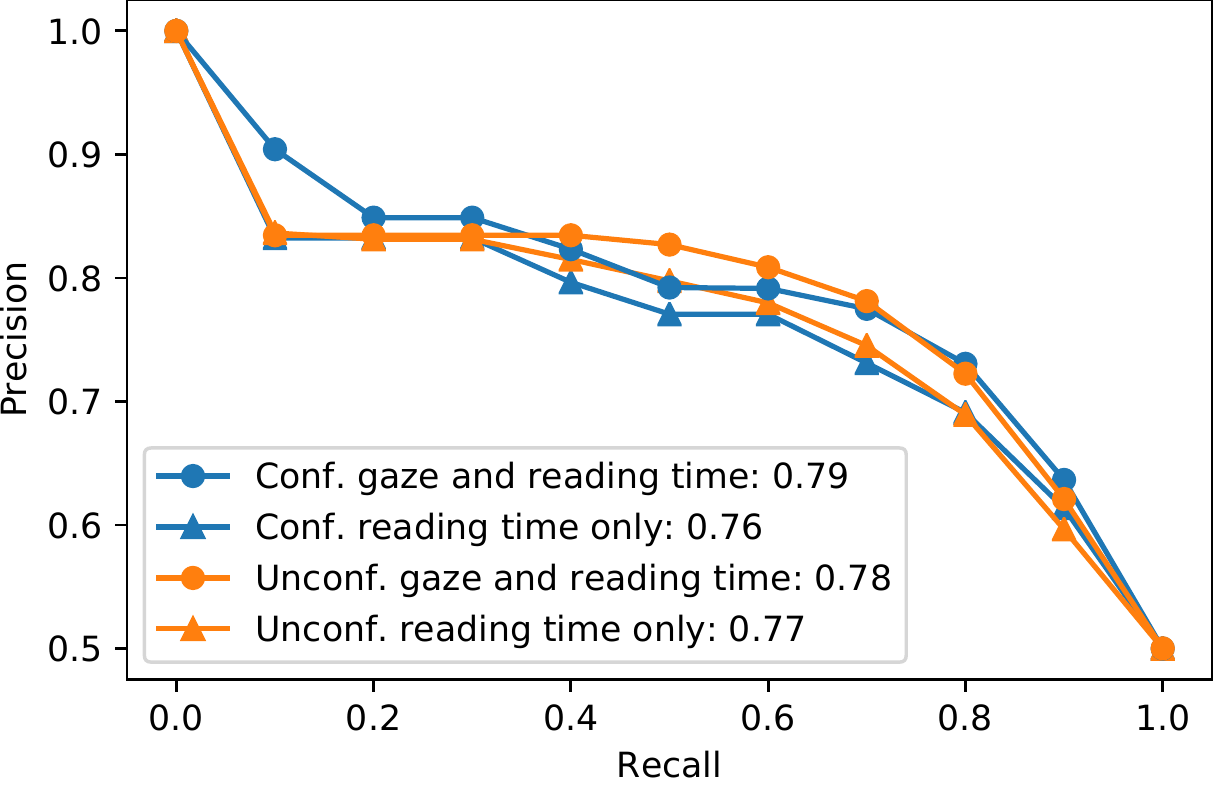}
    \caption{11-point precision-recall curves and average precisions of self-confidence estimations trained by wild data}
    \label{fig:wild_estimation}
\end{figure}

\subsection{Confidence Estimation Results in the Wild}

We utilized the recorded data for training the estimator and Figure~\ref{fig:wild_estimation} shows the estimation results. As same as the laboratory study, our approach could detect confidence and unconfidence relatively better than the estimator using only reading-time. \textit{f1: fixation count on choices}, \textit{f8: minimum fixation duration on choices}, \textit{f12: minimum fixation duration on question}, \textit{f29: reading-time}, and \textit{f30: correctness of the answer} were selected as features.

\begin{figure}[t]
    \centering
    \hspace{1mm}
    \subfloat[Laboratory dataset]{
    \includegraphics[clip,width=0.9\columnwidth]{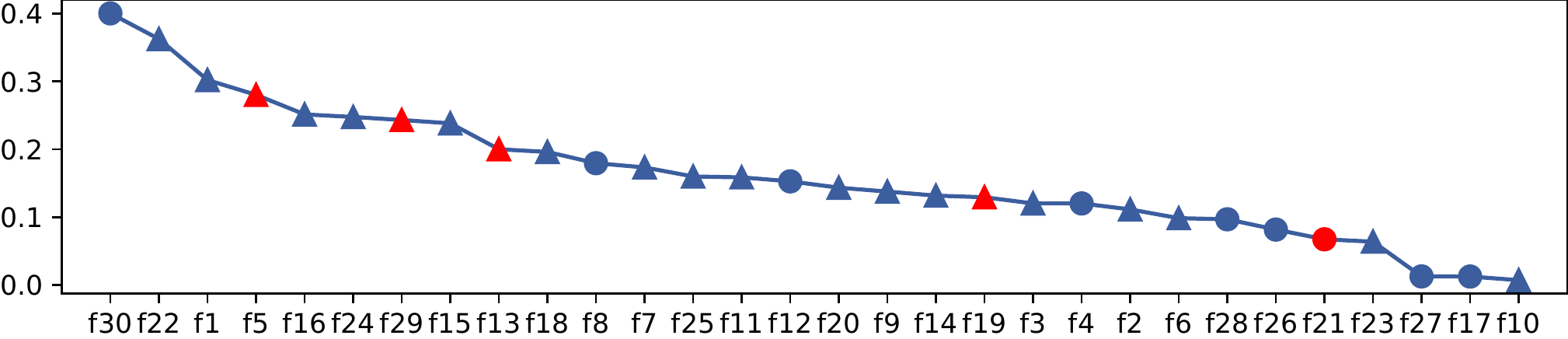}
    \label{figure:correlations_lab}
    }
    \newline
    \subfloat[Wild dataset]{
    \includegraphics[clip,width=0.9\columnwidth]{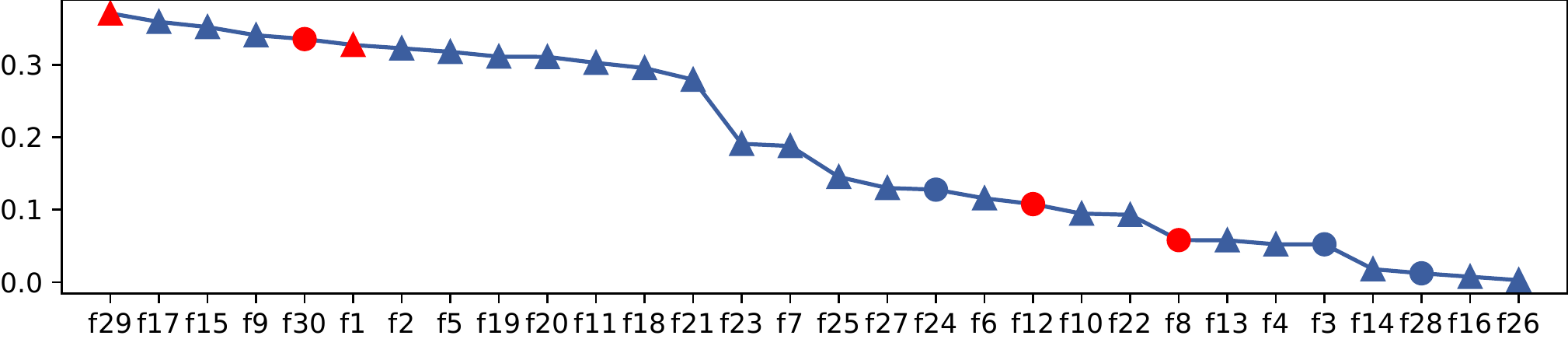}
    \label{figure:correlations_wild}
    }
    \caption[Pearson correlations between self-confidence and each feature]{Pearson correlations between self-confidence and each feature. Features selected by the forward stepwise are highlighted as red color. (circle: positive, triangle: negative correlation; sorted by the absolute value)}
    \label{fig:correlations}
\end{figure}

\subsection{Effective Features}
Figure~\ref{fig:correlations} shows a list of features selected on the laboratory dataset (the first study) and wild dataset (the third study). In both conditions, \textit{f29: reading-time} has a negative correlation with self-confidence, and was selected as a feature. 
Most of the calculated features are negatively correlated with self-confidence. This is because the longer a learner takes time to consider, the more fixations and saccades are observed. Interestingly, a feature that is highly correlated with self-confidence is not necessarily selected in a classifier. Furthermore, a feature that is not correlated individually can play an important role in combining other features.

\subsection{The Number of Training Samples}
\begin{figure}[t]
    \begin{center}
    \hspace{-6mm}
    \includegraphics[clip,width=0.85\hsize]{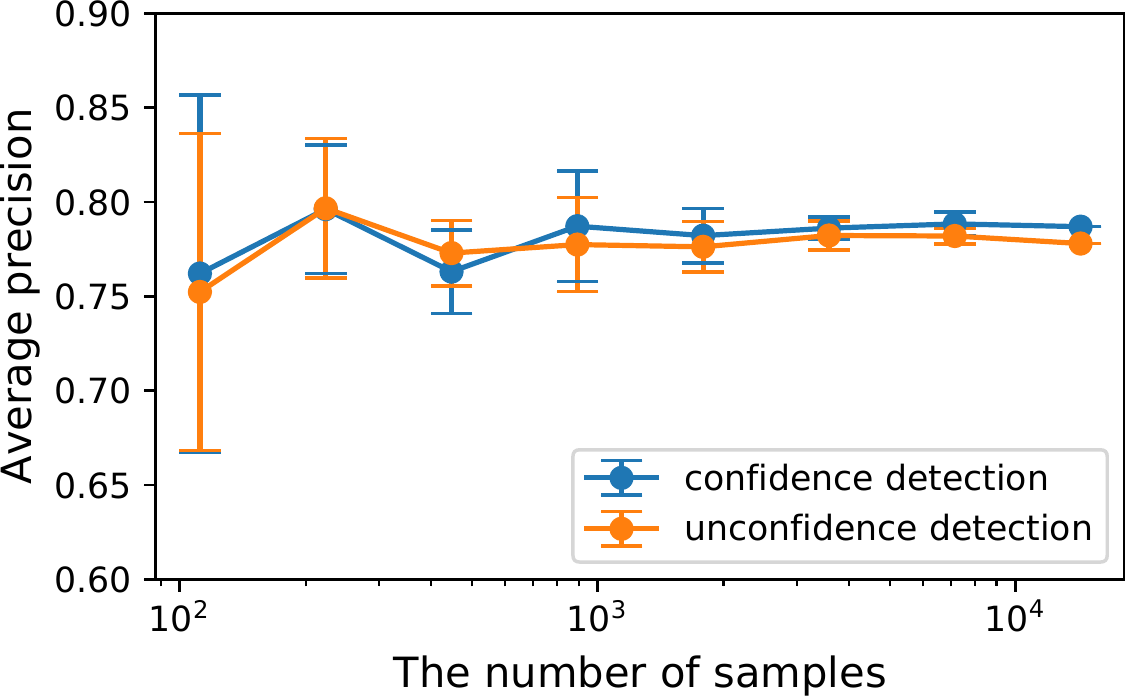}
    \caption{Average precisions on different number of samples randomly selected from the wild dataset.}
    \label{fig:n_data}
    \end{center}
\end{figure}

Figure~\ref{fig:n_data} shows the relation between the number of training samples and the performance. Average precisions increased till the number of training samples reached 200. Increments more than 200 did not contribute to the improvement, but the more training samples we had, the less standard deviation of the result was obtained.

\section{Discussion}

Studies 1 and 3 have given us interesting findings to improve the system. In the first study, we evaluated our gaze-based self-confidence estimation on MCQ. The combination of gaze features and the reading-time could estimate self-confidence better than the estimation by reading-time only (average precisions: 0.81\,\% and 0.79\,\%). One possible reason for the weaker contribution of gaze features compared to the previous report~\cite{yamada2017} is that we aim to develop a system that starts with a user-independent estimation, although individual learners have their own characteristics of eye movements. Our system has a function to collect feedback to the estimation results by learners (see Figure~\ref{fig:overview_report}), and the personalization of the estimation remains for future work.

The third study demonstrated that our self-confidence estimation works in the wild condition such as a real classroom environment, where the system can not be frequently calibrated. Instead of utilizing self-calibration approaches~\cite{huang2016building, santini2017calibme}, calculating features from relative-position based AOIs performed enough in our use case. The number of training samples seems not to be an important matter in this task. Rather than collecting similar answers, recording solving behaviors on varied types and levels of questions with short and long reading-times should improve the estimation.

Another limitation of our studies is that the characteristics of questions in the two datasets were different. Since we could not control the difficulty level of questions in the wild dataset (the third study), the questions seem to be easy for participants, and there are more correct answers than incorrect answers.

The research mentioned in Section~\ref{section:related} attempts are mainly focused on their contexts and parts, and thus it is hard to find the evaluation as a whole system in an end-to-end manner. For example, this means that little has been known how accurate the estimation should be to achieve the goal, which is, in our case, to improve the quality of knowledge. Our second study indicated that questions answered with vague knowledge tend to be forgotten compared to knowledge with confidence (decreased by 16\,\%), and our confidence-based feedback avoided the drop.

An important issue is whether it is still meaningful to give feedback based on the noisy estimation of self-confidence. In order to establish a system that improves the learners' performance, the end-to-end viewpoint must be incorporated into the evaluation. Although there is still room for improvement in our self-confidence estimation, we observed improved learning performances.

\section{Conclusion}

We have proposed \textit{Confidence-Aware Learning Assistant (CoALA)}, which estimates self-confidence on MCQ by analyzing eye movements and generates a report suggesting which question should be reviewed. The self-confidence estimation algorithm was evaluated in the laboratory and the wild condition. By utilizing a pre-trained estimator on the laboratory dataset, we conducted a user-study of the review feedback. Our end-to-end confidence-based review increased correct answer rates by 14\,\% for unconfident correct answers and 17\,\% for confident incorrect answers compared to a controlled condition. By visualizing transitions of correctness and reported self-confidence in a pre-test and a post-test, we observed that the quality of knowledge was increased. We conclude that CoALA is helpful for learners.

In future work, we will apply our method to different kinds of subjects involving Mathematics, Science, Society, etc. We expect a successful estimation of self-confidence in an MCQ, which a student can answer just by looking at a display and thinking about a question. Moreover, we aim to apply our method to questions that do not include choices. In this work, designing AOIs for a question and each choice has been related to obtaining some effective features. We need to find new features to solve this problem.

\bibliographystyle{unsrt}
\bibliography{main}

\EOD

\end{document}